\documentclass[aps,showpacs, preprint]{revtex4}
\usepackage[dvips]{graphicx}
\usepackage{color}
\usepackage{amssymb}
\usepackage{amsmath}
\usepackage{units}
\usepackage[latin1]{inputenc} 

\begin{document}

\title{Brillouin light scattering study of Co$_{2}$Cr$_{0.6}$Fe$_{0.4}$Al and Co$_{2}$FeAl Heusler compounds}

\author{O.~Gaier%
\footnote{Corresponding author: email: gaier@physik.uni-kl.de},
J.~Hamrle,  S.~Trudel, A.~Conca~Parra, B.~Hillebrands}
\affiliation{Fachbereich Physik and Forschungsschwerpunkt OPTIMAS,
Technische Universit\"at Kaiserslautern,
Erwin-Schr\"odinger-Stra\ss e 56, D-67663 Kaiserslautern, Germany}

\author{E.~Arbelo, C.~Herbort, M.~Jourdan}
\affiliation{Institut f\"{u}r Physik, 
Johannes-Gutenberg-Universit\"at Mainz, 
Staudinger Weg 7, D-55099 Mainz, Germany}

\begin{abstract}
The thermal magnonic spectra of Co$_{2}$Cr$_{0.6}$Fe$_{0.4}$Al
(CCFA) and Co$_2$FeAl were investigated using Brillouin light
scattering spectroscopy (BLS). For CCFA, the exchange
constant  $A$ (exchange stiffness $D$) is found to be $0.48\pm0.04$\,$\mu$erg/cm ($
203\pm16$\,meV\,\AA$^2$), while for Co$_2$FeAl the corresponding values of 
 $1.55\pm0.05$\,$\mu$erg/cm ($370\pm10$\,meV\,\AA$^2$) were found.
The observed asymmetry in the BLS spectra between the Stokes and
anti-Stokes frequencies was assigned to an interplay between
the asymmetrical profiles of hybridized Damon-Esbach and perpendicular 
standing spin-wave modes, combined
with the optical sensitivity of the BLS signal to the upper side of the CCFA
or Co$_2$FeAl film.

\end{abstract}


\maketitle

\section{Introduction}
Studies of ferromagnetic (FM) half-metals are mainly driven by
the possible applications of such materials in spintronic devices as a potential
source of a 100\% polarized spin current. Some Heusler alloys are
promising candidates due to their high Curie temperatures \cite{fes06}. For
example, it was recently proven that spin injection from
Co$_2$FeSi into the semiconductor (Al,Ga)As can 
be achieved with a 50\% efficiency \cite{ram08}.

Amongst the different Heusler systems studied in recent years, the
compound Co$_2$Cr$_{0.6}$Fe$_{0.4}$Al (CCFA) has attracted
significant experimental
\cite{fel03,hir05,con07a,wur07, jou07,kse08} and theoretical
\cite{miu04,ant05, wur06} attention. CCFA is an interesting
candidate for spintronics applications
due to its high Curie temperature of \unit[760]{K}
\cite{fel03} and its high value of volume magnetization of
$\approx$\unit[3]{$\mu_B$} per formula unit \cite{con07a,ino06} at
\unit[5]{K} (the theoretical value is \unit[3.8]{$\mu_B$} per
formula unit \cite{gal02}). 
Using spin resolved photoemission, the spin polarization of CCFA at the Fermi level was found to be 45\%  
at room temperature  \cite{cin07}. 
CCFA-based spin-dependent transport devices showing large magnetoresistance effects were recently reported. For example, 
in the simple CCFA/Cu/Co$_{90}$Fe$_{10}$ trilayers
a large giant magnetoresistance of 6.8\% at room temperature (RT)
\cite{kel05} was found.  Tunnelling magnetoresistance ratios (TMR) of 52\% at
RT and 83\% at 5K were reported for the 
CCFA/AlO$_x$/Co$_{75}$Fe$_{25}$ magnetic tunnel junction (MTJ) \cite{ino06}.
In the epitaxial  CCFA/MgO/Co$_{50}$Fe$_{50}$ MTJ structure, TMR ratios of 109\% at RT and 317\% at 4.2\,K \cite{mur07} were presented, whereas for the  
epitaxial CCFA/MgO/CCFA structure 
TMR ratios of 60\% at RT and 238\% at 4.2 K are reported \cite{mar07}.

Thin Co$_2$FeAl films and MTJ structures containing the Co$_2$FeAl
electrode have been a subject of extensive studies as well
\cite{hir05b,hir05c,ino06,tez06,tak08}, even though initial band
structure calculations predicted a much lower degree of spin
polarization compared to  CCFA \cite{miu04}. 
However, recent band structure calculations by Felser \textit{et al.} have predicted \unit[100]{\%} spin polarization at the Fermi level for bulk Co$_2$FeAl \cite{wur06,kan07}.
The TMR
ratios reported in Refs.~\cite{hir05b,hir05c,ino06,tez06,tak08} are
about \unit[50]{\%} at room temperature and are comparable to
those reported for CCFA.

In this article we report on our study of the thermal spectrum of spin waves in CCFA and
Co$_2$FeAl films using Brillouin light scattering (BLS) spectroscopy.
From our BLS spectra we have determined the values of the exchange
constants for CCFA and Co$_2$FeAl.

\section{Sample preparation and characterization}

The CCFA structure under investigation is an
Al(\unit{2.5}{nm})/\-CCFA(\unit{80}{nm})/\-Cr(\unit{8}{nm})/\-MgO(100)
epitaxial structure \cite{con07a, ham06}. The buffer layers were
deposited by electron beam evaporation onto a single-crystalline
MgO(001) substrate, while the epitaxial CCFA films were
subsequently deposited by dc magnetron sputtering. A more detailed
description of the sample preparation can be found
elsewhere \cite{con07a}.  The films grow with the B2 structure, as
there is full disorder between the Cr and Al positions, but order
on the Co positions  \cite{con07a}. A volume magnetization of
$\mu_\mathrm{CCFA}\approx2.5\mu_B$ per f.u. (formula unit) (i.e.\
$M_s=490\,$emu/cm$^3$) was measured at \unit[300]{K} by SQUID magnetometry. 4-circle x-ray diffraction (XRD)
scans yielded lattice parameters \unit[$a=0.570\pm0.005$]{nm} and
\unit[$b = c = 0.583\pm0.012$]{nm}, where the $a$-axis is
perpendicular to the sample surface, and $b$ and $c$ are in-plane
axes \cite{ham06}. The magnetic reversal of the structure is also described in Ref.~\cite{ham06}.

The Co$_2$FeAl sample under investigation consists of an 80~nm
thick Co$_2$FeAl layer which was epitaxially grown on a
single-crystalline MgO(001) substrate covered with a 10~nm thick
MgO buffer layer. Magnetron sputtering was employed for the deposition of both the MgO buffer and
the Co$_2$FeAl layer. A
post-growth anneal at 550$^\circ$C provided a 
Co$_2$FeAl film with the B2 structure, as confirmed by XRD measurements. A 3~nm thick AlO$_x$
capping layer was deposited on top of the structure  to prevent sample oxidation.
The sample exhibits a saturation magnetization of 4.66~$\mu_B$/f.u.\ measured by SQUID magnetometry at room temperature. This is in
a good agreement with the previously reported value of
4.96~$\mu_B$/f.u.\ determined at 4.2~K \cite{bus83}.

All BLS measurements presented in this article 
were performed using a diode pumped, frequency doubled Nd:YVO$_4$ laser with a wavelength of $\lambda=532$~nm as a light source. 
Unless specified otherwise, the light impinges on the sample at an angle of incidence of $\varphi=45^\circ$, corresponding to a
transferred wave vector of detected magnons 
$q_\parallel=4\pi/\lambda\sin\varphi=1.67\cdot10^{5}$\,cm$^{-1}$. 
The external magnetic field~$H$ was applied in the so-called magnetostatic
surface mode geometry, wherein $H$ is applied in the plane of the sample, and is
perpendicular to the plane of light incidence (i.e.,
$\vec{H}\perp\vec{q}_\parallel$). 
A more detailed description of the BLS setup used in this work can be found in Refs.~\cite{moc87, hil99}.

\section{Experimental results}

\subsection{CCFA}

\begin{figure}
\includegraphics[width=0.6\textwidth]{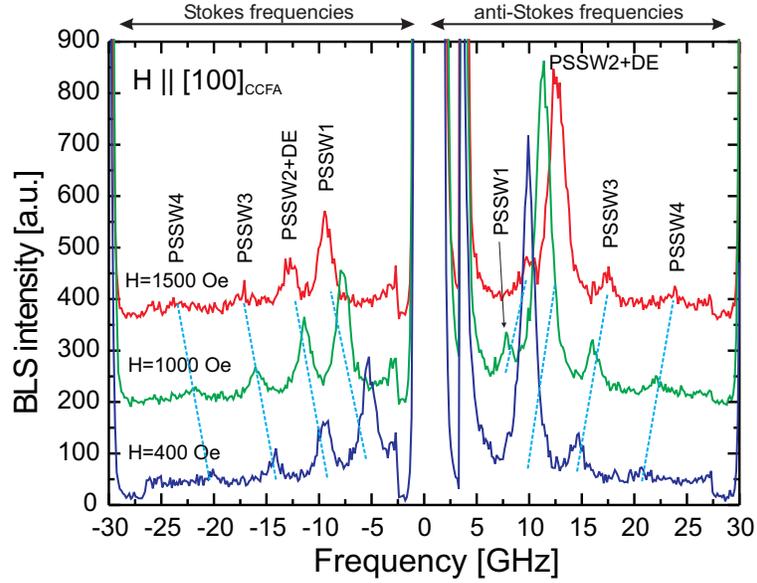}
\caption{%
\label{ccfa_exam}%
(color online) Selected BLS spectra measured on the CCFA(80\,nm) film in an
external magnetic field of $H=400$, 1000 and 1500\,Oe. Negative
frequencies are related to Stokes processes (creation of magnons),
whereas positive frequencies are related to anti-Stokes processes
(annihilation of magnons). The dashed lines are  guides to the eye.}
\end{figure}

Typical BLS spectra measured on the investigated CCFA film are
presented in Fig.~\ref{ccfa_exam} for several values of the
external magnetic field. 
As is clearly visible, the positions of the peaks
in both the Stokes 
(creation of magnons, negative frequencies in BLS spectra) 
and anti-Stokes 
(annihilation of magnons, positive frequencies in BLS spectra)
parts 
of the spectrum move to
higher values with an increasing magnetic field. This field dependence is
evidence of the magnonic nature of the observed peaks, whereas the position of a phononic peak is not expected to be field dependent.
The observed peaks correspond to the
Damon-Eshbach (DE) mode and to the perpendicular standing
spin-wave (PSSW) modes.  As will be discussed 
below, one of the observed peaks in the BLS spectra results from the
hybridization of the DE and the second PSSW mode. Therefore, this
mode will be referred as PSSW2+DE in the following discussion. The other observed
peaks are pure PSSW modes.

\begin{figure}
\includegraphics[width=0.9\textwidth]{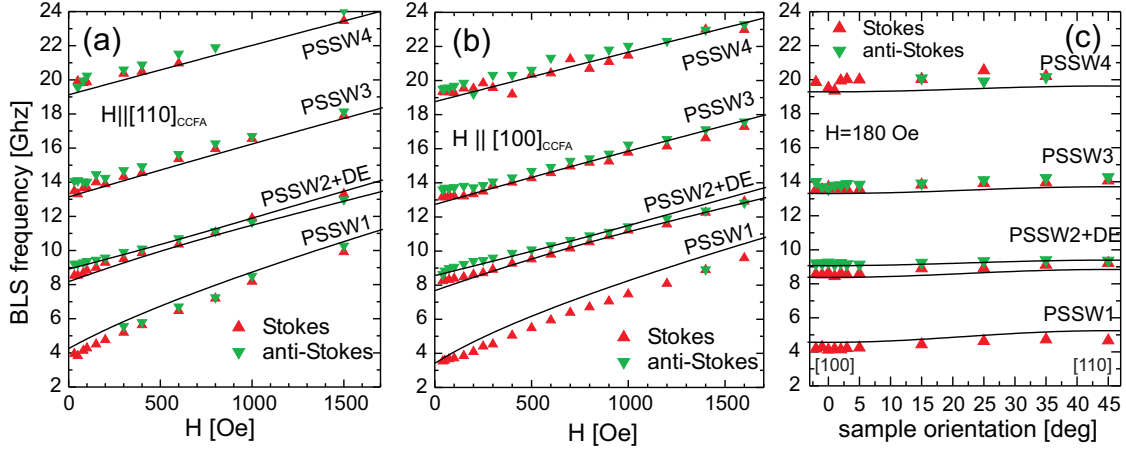}%
\caption{%
\label{ccfa_fh}%
(color online) (symbols) (a)(b) Dependence of the experimental BLS frequencies on
the external magnetic field for $H$ applied along the 
[110]$_\mathrm{CCFA}$ (a) and the  [100]$_\mathrm{CCFA}$ axis (b). (c)
Dependence of the BLS frequencies on the sample orientation, i.e.
the angle between the [100]$_\mathrm{CCFA}$ axis and the in-plane applied
magnetic field. Triangles-up (down) refer to Stokes (anti-Stokes)
frequencies of the BLS spectra. In all three panels, solid lines are the calculated spin-wave
frequencies in CCFA(80\,nm) using the exchange constant
$A=0.48$\,$\mu$erg/cm, saturation magnetization
$M_S=520$\,emu/cm$^3$, Land\'e $g$-factor $g=1.9$, transferred
wave vector $q_\parallel=1.67\cdot10^{5}$\,cm$^{-1}$ and cubic
volume anisotropy $K_1=-20$\,kerg/cm$^3$.}
\end{figure}

Figures~\ref{ccfa_fh}(a,b) provide the dependence of the observed
spin wave frequencies on the value of the external magnetic field
which was applied along the [110] (Fig.~\ref{ccfa_fh}(a)) and the
[100] (Fig.~\ref{ccfa_fh}(b)) directions of the CCFA film. The field
dependencies are nearly identical for both orientations of the
external magnetic field. However, the observed spin wave
frequencies are about 1\,GHz smaller when $H$ is aligned
parallel to the [100]$_\mathrm{CCFA}$ direction, compared to when the field is applied along the [110] direction. This is also
illustrated in Fig.~\ref{ccfa_fh}(c), where the dependence of the
observed spin wave frequencies on the in-plane sample orientation (i.e., the angle between [100] direction and the applied magnetic field)
is presented. 
The lower frequency observed when the field is along  the [100]$_\mathrm{CCFA}$ direction
 indicates that this direction is an in-plane magnetically hard axis \cite{BLS}, in agreement with previous MOKE
investigations \cite{ham06}. 

The solid lines in
Figures~\ref{ccfa_fh}(a--c) are a fit to the experimental data
using a theoretical model described in Ref.~\cite{hil90}. The
found parameters are: exchange constant
$A=0.48\pm0.04$\,$\mu$erg/cm, cubic volume anisotropy
$K_1=-20\pm10$\,kerg/cm$^3$, Land\'e $g$-factor $g=1.9\pm0.1$ and saturation magnetization $M_S=520\pm20$\,emu/cm$^3$. Note that each of the fitted parameter is determined rather independently by a particular experimental dependence or feature, making the fit reliable and providing results with a relatively small error. Namely, 
$M_S$ is obtained from the frequency of the DE mode (here PSSW2+DE mode), and agrees well with the SQUID value  $M_S=490$\,emu/cm$^3$
\cite{con07a}. Land\'e $g$-factor is provided by the slope of BLS frequency on the external field. $K_1$ is determined from the variation of BLS frequencies on the sample orientation (Fig.~\ref{ccfa_fh}(c)). Finally, the exchange constant $A$ is determined by frequencies of PSSW modes. Note, that the corresponding exchange stiffness of CCFA is 
$D=2Ag\gamma_0\hbar / M_S=203\pm16$\,meV\,\AA$^2$, 
where $\hbar$ and $\gamma=g \gamma_0$ are the reduced Planck constant and the gyromagnetic ratio, respectively. 

For the PSSW2+DE mode we observe a large (between 0.5--1\,GHz)
splitting between the Stokes 
($\blacktriangle$) 
and the anti-Stokes 
($\blacktriangledown$)
frequencies
(Figs.~\ref{ccfa_fh}(a,b)). This splitting is particularly
pronounced for small values of the external field in the range of
100--400\,Oe. A careful investigation of Figs.~\ref{ccfa_fh}(a,b)
further reveals that the lower frequency component of the split
PSSW2+DE mode are observed only in the Stokes part of the BLS
spectrum, whereas the higher frequency component appears only in the
anti-Stokes part. Moreover, the slope d$f$/d$H$ is slightly
different for frequencies determined from the Stokes and 
anti-Stokes parts of the BLS spectra. In the case of a DE
mode, such an asymmetry usually indicates different pinning
conditions of the dynamic magnetization
on each interface
\footnote{The dynamic magnetization is the difference between the static
magnetization $\vec{M}_0$ and its  instantaneous value $\vec{M}(t)$:
$\vec{m}(t)=\vec{M}(t)-\vec{M}_0$.}.
However, in our particular case, the observed
asymmetry is related to the hybridization (also called mode
repulsion) of the DE and the PSSW2 modes, which is elaborated 
in detail in the Discussion section.

\subsection{C\lowercase{o}$_2$F\lowercase{e}A\lowercase{l}}

\begin{figure}
\begin{center}
\includegraphics[width=1.0\textwidth]{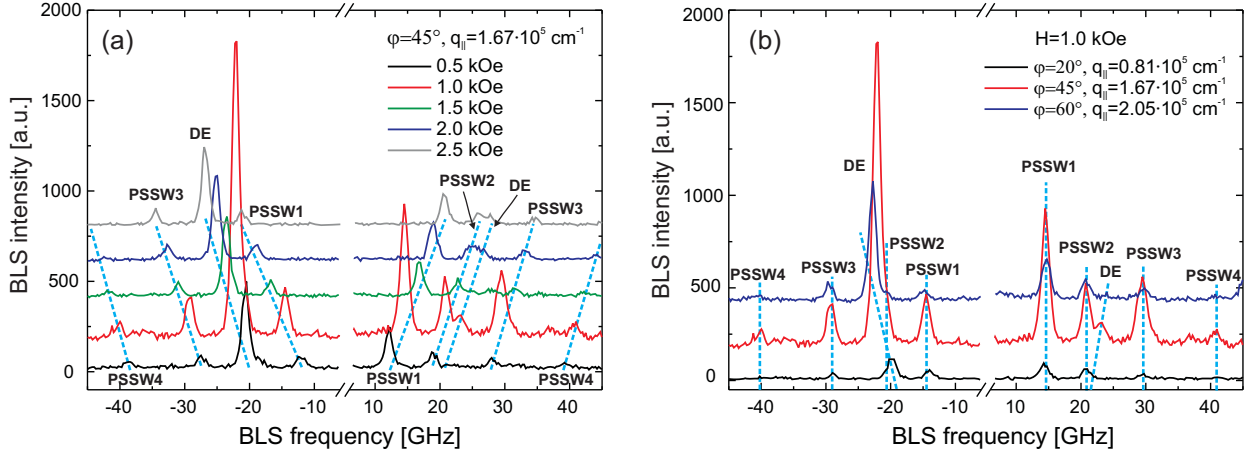}
\end{center}
\caption{%
\label{f:cfa10_spectra}%
(color online) (a) BLS spectra of Co$_2$FeAl recorded at a
transferred wave vector $q_\|=1.67\cdot10^5$~cm$^{-1}$ and
different magnetic fields. The change of peak positions upon the
variation of the magnetic field reveals the magnonic origin of the
peaks. (b) BLS spectra of Co$_2$FeAl recorded in an external
applied field $H$ of 1~kOe and different incidence angles $\varphi$. In (a-b), the dashed lines are guides to the eye. 
The DE mode exhibits a strong dependence on $\varphi$, whereas the
frequency of the PSSW modes does not significantly change with
$\varphi$. A clear distinction between the DE and PSSW modes
thus becomes  possible. For the assignment of the standing spin
wave peaks to different modes see the text.}
\end{figure}

The BLS spectra collected from the Co$_2$FeAl sample are shown in
Fig.~\ref{f:cfa10_spectra}. In (a) spectra recorded at a
transferred wave vector $q_\|=1.67\cdot10^5$\,cm$^{-1}$ and
different values of the external magnetic field are displayed,
while in (b) BLS spectra measured at $H=1$\,kOe and at different 
$q_\|$ are presented. As is clearly visible from
Fig.~\ref{f:cfa10_spectra}(a), the positions of the peaks
move to higher values as the  magnetic field increases, confirming 
the magnonic origin of the observed peaks.

Performing BLS measurements at different angles of incidence
$\varphi$ (i.e. at different values of $q_\|$), allows for an
unambiguous separation of the dipole dominated magnetostatic
surface wave (the Damon-Eshbach mode), from the
exchange dominated perpendicular standing spin waves (PSSW). In
contrast to PSSW modes, the frequency of the DE mode exhibits a
much stronger dependence on the in-plane direction of the wave vector. 
Therefore it provides
substantial shifts in the collected spectra upon the variation of
$\varphi$ (and thus $q_\|$), while the spectral positions of the PSSW modes are not significantly affected by this variation. Therefore the peak originating from the DE mode excitation
can be easily identified. (Fig.~\ref{f:cfa10_spectra}(b)).

\begin{figure}
\begin{center}
\includegraphics[width=0.9\textwidth]{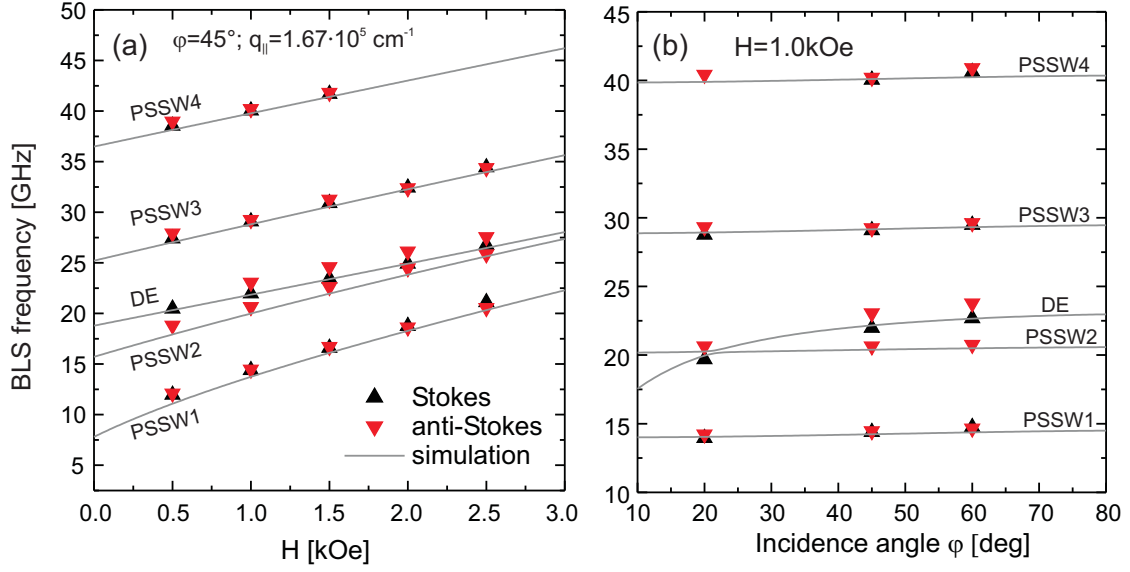}
\end{center}
\caption{%
\label{f:cfa10_simulation}%
(color online) (symbols) Experimental BLS
frequencies of Co$_2$FeAl, corresponding to the peak positions in
the BLS spectra presented in Fig.~\ref{f:cfa10_spectra}(a) and
(b). (solid lines) Simulations for the exchange constant
$A=1.55$\,$\mu$erg/cm, saturation magnetization
$M_S=1027$\,emu/cm$^3$ and Land\'e $g$-factor $g=2.1$. (a) shows
the dependence of BLS frequencies on the external magnetic field,
whereas (b) demonstrates the dependence of BLS frequencies on the
angle of incidence $\varphi$.}
\end{figure}

The extracted peak
positions as a function of  external magnetic field and 
incidence angle are presented by symbols in Fig.~\ref{f:cfa10_simulation}(a) and (b), respectively. 
The solid lines present the simulation from which
the following parameters were determined: exchange constant
$A=1.55\pm0.05$\,$\mu$erg/cm, saturation magnetization
$M_S=1027\pm10$\,emu/cm$^3$ and Land\'e $g$-factor $g=2.1\pm0.1$. The
corresponding exchange stiffness is $D=370\pm10$\,meV\,\AA$^2$.
The value of the cubic volume anisotropy $K_1$ was not determined and
hence its value is assumed to be zero in the simulations. 
Note that including $K_1$ into the model only slightly changes the frequency of the spin waves (by about 1 GHz). Therefore, setting $K_1$ to zero in the model used does not detrimentally affect the determination of $M_S$ and $A$.

Note that in the Stokes part of the spectrum (i.e.\ negative
frequencies) shown in Fig.~\ref{f:cfa10_spectra}(a), the PSSW2 mode is
difficult to recognize. This is due to a relatively small spacing
between the PSSW2 and DE modes, as well as the much stronger intensity
of the adjacent DE mode. 
In the anti-Stokes region of the spectrum, the DE mode has intensities that are comparable to, or lesser than the PSSW2 mode.
The origin of
this asymmetry of the DE mode intensities will be discussed in the
following.

\section{Discussion}

The values of exchange constant (exchange
stiffness) of CCFA was found to be $A=0.48\pm0.04$\,$\mu$erg/cm ($D=
203\pm16$\,meV\,\AA$^2$), whereas the corresponding value for 
Co$_2$FeAl was found
to be $A=1.55\pm0.05$\,$\mu$erg/cm ($D=370\pm10$\,meV\,\AA$^2$). 
 This exchange constant found for Co$_2$FeAl is smaller than both the exchange constants of
bcc Fe ($A_\mathrm{Fe}=$\unit[2.0]{$\mu$erg/cm},
$D_\mathrm{Fe}=280$\,meV\,\AA$^2$ \cite{hil90,pau82a,shi68}) and of bcc Co
($A_\mathrm{Co}=2.12$\,$\mu$erg/cm,
$D_\mathrm{Co}=430$\,meV\,\AA$^2$ \cite{liu96}). However, the exchange stiffness of Co$_2$FeAl is between the exchange stiffness of bcc Fe and bcc Co.
On the other
hand, the value of the exchange constant (exchange stiffness) of CCFA is very small, being only one
third (one half) of the exchange constant (exchange stiffness) of Co$_2$FeAl, and about one half of the exchange constant or exchange stiffness 
of fcc Ni ($A_\mathrm{Ni}=0.85$\,erg/cm,
$D_\mathrm{Ni}=420$\,meV\,\AA$^2$). The large discrepancy between the values of exchange determined for Co$_2$FeAl and CCFA demonstrates how these can vary substantially in Co$_2$-based Heusler compounds, even for compounds with the same ordering (B2) and similar compositions.

The value of the cubic volume anisotropy $K_1$ in CCFA is
found to be $-20\pm10$\,kerg/cm$^3$, being about 24~times smaller than
$K_1$ of bcc Fe ($K_{1,\mathrm{Fe}}=480$\,kerg/cm$^3$
\cite{ram58}). However, it is roughly of the order of bcc Co 
($K_{1,\mathrm{Co}}=0\pm10$\,kerg/cm$^3$ \cite{liu96}). The small
value  of $K_1$ in CCFA found here is consistent with previous investigations showing
that the magnetic anisotropy is weak in Heusler compounds \cite{gai08,liu07}, reflecting the small anisotropy of the spin-orbit coupling in these materials \cite{siegmann}.

\begin{figure}
\includegraphics[width=0.8\textwidth]{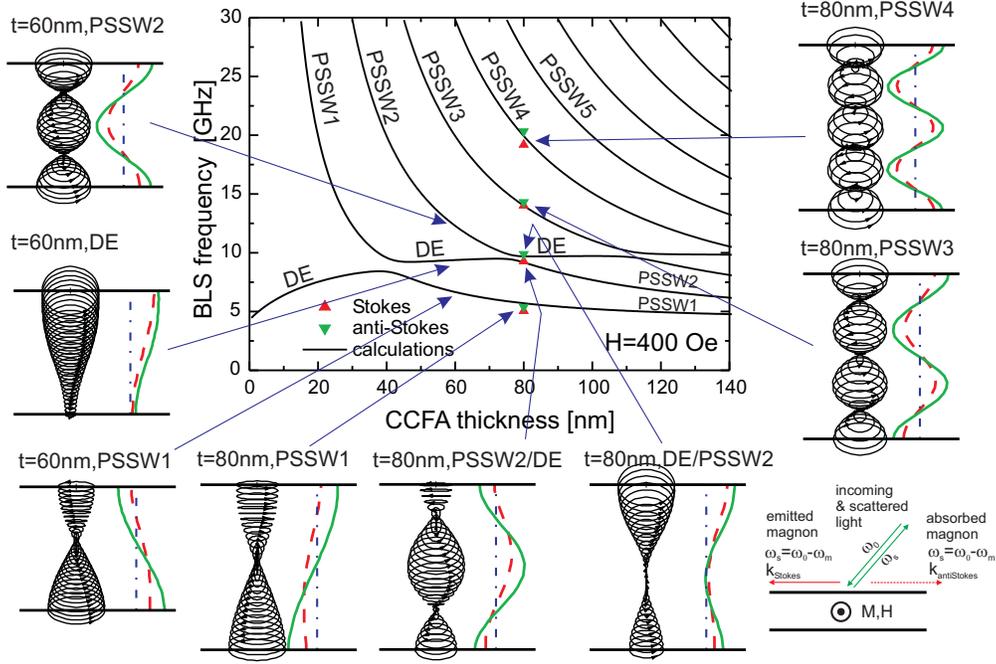}
\caption{%
\label{ccfa_tdep}%
(color online) Calculated dependence of spin-wave frequencies on the thickness of
the CCFA film for $H=400$\,Oe, $H\parallel [100]_\mathrm{CCFA}$
and a transferred spinwave wavevector
$q_\|=1.67\cdot10^5$\,cm$^{-1}$. Symbols (solid line) represent
experimental data (simulations). The parameters used
in the simulations are the same as in Fig.~\protect\ref{ccfa_fh}.
The graph is surrounded by calculated profiles of spin-wave modes
over the film thickness for $t=60$ and 80\,nm. The left part of
each profile image shows the trajectory of magnetization when
looking onto the magnetization vector. The right part shows the
corresponding profile of the amplitude of the dynamic
magnetization in the out-of plane (red dashed line) and the in-plane
(green solid line) directions, respectively. The sketch in the
bottom-right corner shows the used geometry.}
\end{figure}

To explain the aforementioned asymmetries between the Stokes 
and anti-Stokes peak positions in the BLS spectra,
we calculated the depth profiles of the dynamic magnetization for CCFA and Co$_2$FeAl films, using the model presented in
Ref.~\cite{hil90}. The results for  CCFA are presented
in Fig.~\ref{ccfa_tdep} for CCFA thicknesses $t=60$ and 80\,nm, along with the
calculated dependence of the spin wave frequencies on the CCFA
thickness. This figure shows that in the vicinity of $t=80$\,nm,
i.e. the thickness of the investigated CCFA film, a crossing
between the DE and the PSSW modes occurs, resulting in a
hybridization of these two modes. Due to the hybridization, a gap
of 0.3\,GHz is created, which corresponds to the splitting between
the Stokes and anti-Stokes frequencies.

The hybridization is also visible in the calculated depth profiles
of the dynamic magnetization in Fig.~\ref{ccfa_tdep}. The left part of each profile image
shows the trajectories of the magnetization vector along the film depth (the trajectory at a given depth is
 an ellipse). In the corresponding right part of each image the
profiles of the in-plane and out-of-plane components of the
dynamic magnetization are shown, denoted by  green solid and red
dashed lines, respectively. Note that due to the shape anisotropy
of the CCFA film, the amplitude of the in-plane dynamic magnetization
is roughly twice as large as the amplitude in the out-of plane
direction. 
As demonstrated for
$t=60$\,nm, the DE mode has a large dynamic magnetization nearby
one interface and decays to the second interface. 
The $m$-th order mode of the PSSW 
 has $m$ nodes, (i.e., there is $m$ points in the depth of the
CCFA film where the dynamic magnetization is zero) and is symmetrical with respect to both interfaces \cite{kal86}. For the $80$\,nm
thick film, the profiles of both hybridized modes (denoted as
DE/PSSW2 and PSSW2/DE) have characteristic features of both types of modes, namely they
exhibit a larger dynamic magnetization nearby one of the interfaces (a characteristic feature
of the DE mode) and they support two nodes (a feature of the PSSW2 mode).
A careful investigation of those two modes shows that when the maximum
dynamic amplitude for the DE/PSSW2 mode is at the upper interface,
the PSSW2/DE mode has a larger amplitude at the opposite interface.

The presented depth profiles of the dynamic magnetization shown in 
Fig.~\ref{ccfa_tdep} were calculated for the anti-Stokes modes. For the Stokes modes, the profiles are reversed, i.e. modes that were
bound to the upper interface are now bound to the bottom interface,
and \textit{vice versa}. Since the typical probing depth for 
the laser wavelength used in our BLS experiments is
about 20--30\,nm \cite{ham02, buch07}, the experiments presented here are only
sensitive to the dynamic magnetization nearby the upper interface. Furthermore, the sensitivity to
the out-of-plane magnetization component is about twice as large
compared to in-plane component in the given geometry
($\varphi=45^\circ$). Combining these two points with the
discussed mode profiles shows that the DE/PSSW2 (PSSW2/DE) mode
provides a larger signal in the anti-Stokes (Stokes) part of the
BLS spectra. This explains why different spin wave modes are
observed in the Stokes and anti-Stokes parts of the BLS spectra of the CCFA film.

\begin{figure}
\includegraphics[width=0.8\textwidth]{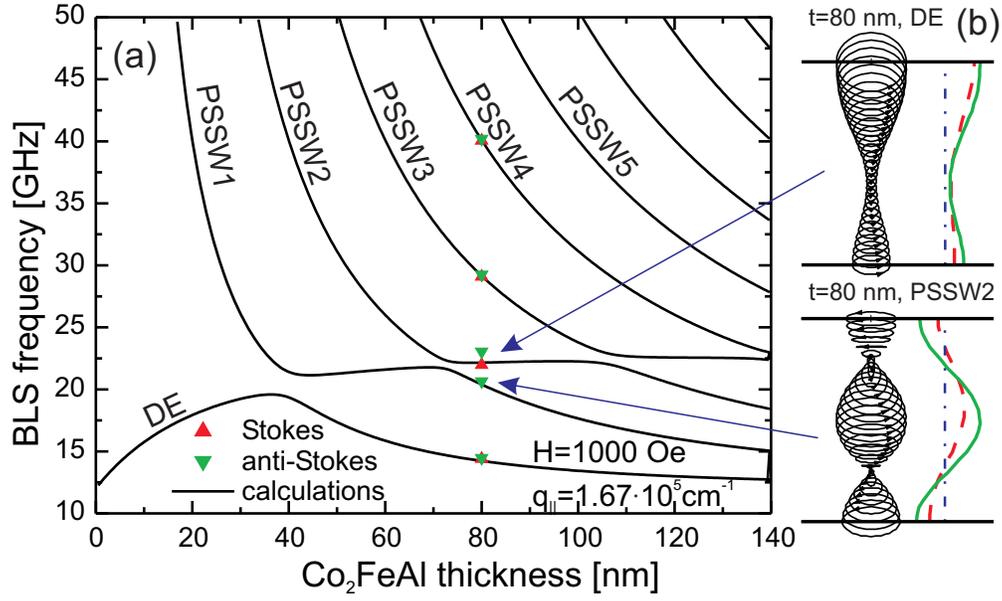}
\caption{%
\label{cfa_tdep}
(color online) (a) Calculated dependence of spin-wave frequencies on the thickness of
the Co$_2$FeAl film for $H=1000$\,Oe and
for spinwave's transferred wavevector
$q_\|=1.67\cdot10^5$\,cm$^{-1}$. Symbols (solid line) represent
experimental data (simulations). The parameters used
in the simulations are the same as in Fig.~\protect\ref{f:cfa10_simulation}.
(b) Profiles of dynamic magnetization for the DE and PSSW2 modes, 
calculated for $t=80$\,nm.}
\end{figure}

There are several analogies  between the mode splitting observed in the CCFA film and the BLS spectra observed in the Co$_2$FeAl film (Fig.~\ref{f:cfa10_simulation}). (i) The DE and PSSW2 modes observed in the Co$_2$FeAl are rather close to each other (about 2\,GHz) (ii) the DE and PSSW2 mode frequencies have different slopes d$f/$d$H$. (iii) In the Stokes part of the BLS spectra, only the DE mode is visible, since the BLS intensity of the DE mode is much larger than the BLS intensity of PSSW2 mode. 
(iv) The dependence of the BLS frequency on the Co$_2$FeAl thickness (presented in Fig.~\ref{cfa_tdep}(a)) is similar to that of CCFA (Fig.~\ref{ccfa_tdep}). In particular, the DE and PSWW2 modes cross each other at a thickness of about 80\,nm in both cases.
(v) The calculated profiles of the dynamic magnetization in Co$_2$FeAl for the DE and PSSW2 modes (shown in Fig.~\ref{cfa_tdep}(b)) demonstrate that they are hybridized. Furthermore, the profiles of those modes are very similar to the profiles of hybridized modes at $t=80$\,nm found in the CCFA film (Fig.~\ref{ccfa_tdep}). 

Therefore, the BLS intensities of each mode are expected to be similar too, which is indeed observed  experimentally.
While the DE and PSSW2 waves are indistinguishable in CCFA due to a small difference in frequency between these two modes, the  frequency difference between these same two modes is larger in the anti-Stokes part of the BLS spectra for Co$_2$FeAl, which allowed us to differentiate them.
As such, in the case of Co$_2$FeAl, we can see that the intensity of the DE mode is smaller than the intensity of the PSSW2 mode on the anti-Stokes side. It confirms our discussion regarding CCFA, showing that the observed mode splitting in CCFA is an interplay between asymmetrical profiles of hybridized modes and the optical depth selectivity of the BLS signal.

\section{Conclusion}

We have investigated thermal spin waves in CCFA and Co$_2$FeAl
using BLS spectroscopy. The values of the exchange constant (exchange
stiffness) of CCFA was found to be $A=0.48\pm0.04$\,$\mu$erg/cm ($D=
203\pm16$\,meV\,\AA$^2$) whereas the exchange of Co$_2$FeAl was found to be 
$A=1.55\pm0.05$\,$\mu$erg/cm ($D=370\pm10$\,meV\,\AA$^2$). 
The found cubic volume magnetic anisotropy of CCFA $K_1=-20\pm10$\,kerg/cm$^3$ is in agreement with small values of $K_1$ found for other Heusler compounds.

The observed asymmetry in BLS spectra between Stokes and
anti-Stokes frequencies was assigned to an interplay between
asymmetrical profiles of hybridized DE and PSSW2 modes, combined
with the optical sensitivity of the BLS signal to the upper side of the CCFA
or Co$_2$FeAl film, due to a limited probing depth.

\section{Acknowledgment}

The project was financially supported by the Research Unit 559
\emph{``New materials with high spin polarization''} funded by the
Deutsche Forschungsgemeinschaft  and by the Stiftung
Rheinland-Pfalz f\"ur Innovation.

\clearpage



\end{document}